\documentclass{article}

\usepackage[nonatbib,preprint]{neurips_data_2024}

\usepackage[utf8]{inputenc} 
\usepackage[T1]{fontenc}    
\usepackage{url}            
\usepackage{booktabs}       
\usepackage{amsfonts}       
\usepackage{nicefrac}       
\usepackage{microtype}      
\usepackage{xcolor}         
\usepackage{subfigure}
\usepackage{multirow}
\usepackage{makecell}
\usepackage{diagbox}
\usepackage{bbding}
\usepackage{amsmath}
\usepackage{wrapfig}
\usepackage{placeins}
\usepackage{graphicx}
\usepackage{float}
\usepackage{subcaption}

\title{VulDetectBench:
	Evaluating the Deep Capability of Vulnerability Detection with\\ Large Language Models}

\author{Yu Liu\textsuperscript{1,4}\footnotemark[2] \qquad Lang Gao\textsuperscript{2}\footnotemark[2]
\qquad Mingxin Yang\textsuperscript{2}\footnotemark[2] 
\\ \textbf{Yu Xie\textsuperscript{3}\footnotemark[1]} 
\qquad \textbf{Ping Chen\textsuperscript{1}} \qquad \textbf{Xiaojin Zhang\textsuperscript{2}} \qquad \textbf{Wei Chen\textsuperscript{5}\footnotemark[1]} \\
\textsuperscript{1}Institute of BigData, Fudan University, Shanghai, China \\
\textsuperscript{2}School of Computer Science, Huazhong University of Science and Technology, China \\
\textsuperscript{3}Purple Mountain Laboratories, Nanjing, China \\
\textsuperscript{4}School of Computer Science, Fudan University, Shanghai, China\\
\textsuperscript{5}School of Software Engineering, Huazhong University of Science and Technology, China \\
}

\begin{document}

\maketitle

\renewcommand{\thefootnote}{\fnsymbol{footnote}} 
\footnotetext[1]{Corresponding authors.}
\footnotetext[2]{Equal Contributions.}
\begin{abstract}
  Large Language Models (LLMs) have training corpora containing large amounts of program code, greatly improving the model's code comprehension and generation capabilities. However, sound comprehensive research on detecting program vulnerabilities, a more specific task related to code, and evaluating the performance of LLMs in this more specialized scenario is still lacking. To address common challenges in vulnerability analysis, our study introduces a new benchmark, VulDetectBench, specifically designed to assess the vulnerability detection capabilities of LLMs. The benchmark comprehensively evaluates LLM's ability to identify, classify, and locate vulnerabilities through five tasks of increasing difficulty. We evaluate the performance of 17 models (both open- and closed-source) and find that while existing models can achieve over 80\% accuracy on tasks related to vulnerability identification and classification, they still fall short on specific, more detailed vulnerability analysis tasks, with less than 30\% accuracy, making it difficult to provide valuable auxiliary information for professional vulnerability mining. Our benchmark effectively evaluates the capabilities of various LLMs at different levels in the specific task of vulnerability detection, providing a foundation for future research and improvements in this critical area of code security. VulDetectBench is publicly available at \url{https://github.com/Sweetaroo/VulDetectBench}.

\end{abstract}

\section{Introduction}

Recent advancements in Large Language Models (LLMs) have demonstrated remarkable capabilities across various domains. The training datasets of some of these LLMs encompass vast quantities of program code, or the models have been fine-tuned on specific code datasets, endowing LLMs with significant abilities\cite{shi2023towards} in code understanding, generation, and summarizing. This development leads to a deeper investigation into whether current LLMs possess the capacity for vulnerability detection within code and the extent of effectiveness they can achieve in this realm. 

Vulnerability analysis is a complex and systematic task, with numerous traditional methods and methods using deep learning have been proposed. Traditional methods focus on static program analysis\cite{beller2016analyzing, mirsky2023vulchecker}and dynamic testing\cite{8835316, coredumpAmericanFuzzy}to identify and locate vulnerabilities in programs, improving the accuracy of analysis results while reducing false positives, and improving analysis efficiency and scalability for large programs have been the focus of research in traditional methods. Previous work based on deep learning approaches \cite{li2018vuldeepecker} creates a dataset comprising programs and their associated vulnerabilities. It applies deep learning techniques to determine the presence of vulnerabilities in a program or predict their specific type. However, due to limitations in dataset size and quality, these methods struggle to handle cases that fall beyond the training data distribution, thus hindering effective application in real-world scenarios \cite{chakraborty2021deep}.

Currently, there has been research into the capabilities of LLMs in identifying and repairing vulnerabilities in code, and the creation of datasets to analyze the ability of LLMs to detect vulnerabilities (i.e. determine whether code has vulnerabilities). However, a benchmark for evaluating the vulnerability detection capabilities of LLMs and a comparative analysis of various known open source and proprietary LLMs against such a benchmark is lacking. Different LLMs, due to their different sizes in terms of parameters, have different capabilities in different subtasks of vulnerability detection. Using a reasonable benchmark to evaluate the performance of different LLMs in these subtasks can provide critical reference information for subsequent research. 


\begin{wrapfigure}{r}{0.5\textwidth}
    \centering
    \includegraphics[width=0.8\linewidth]{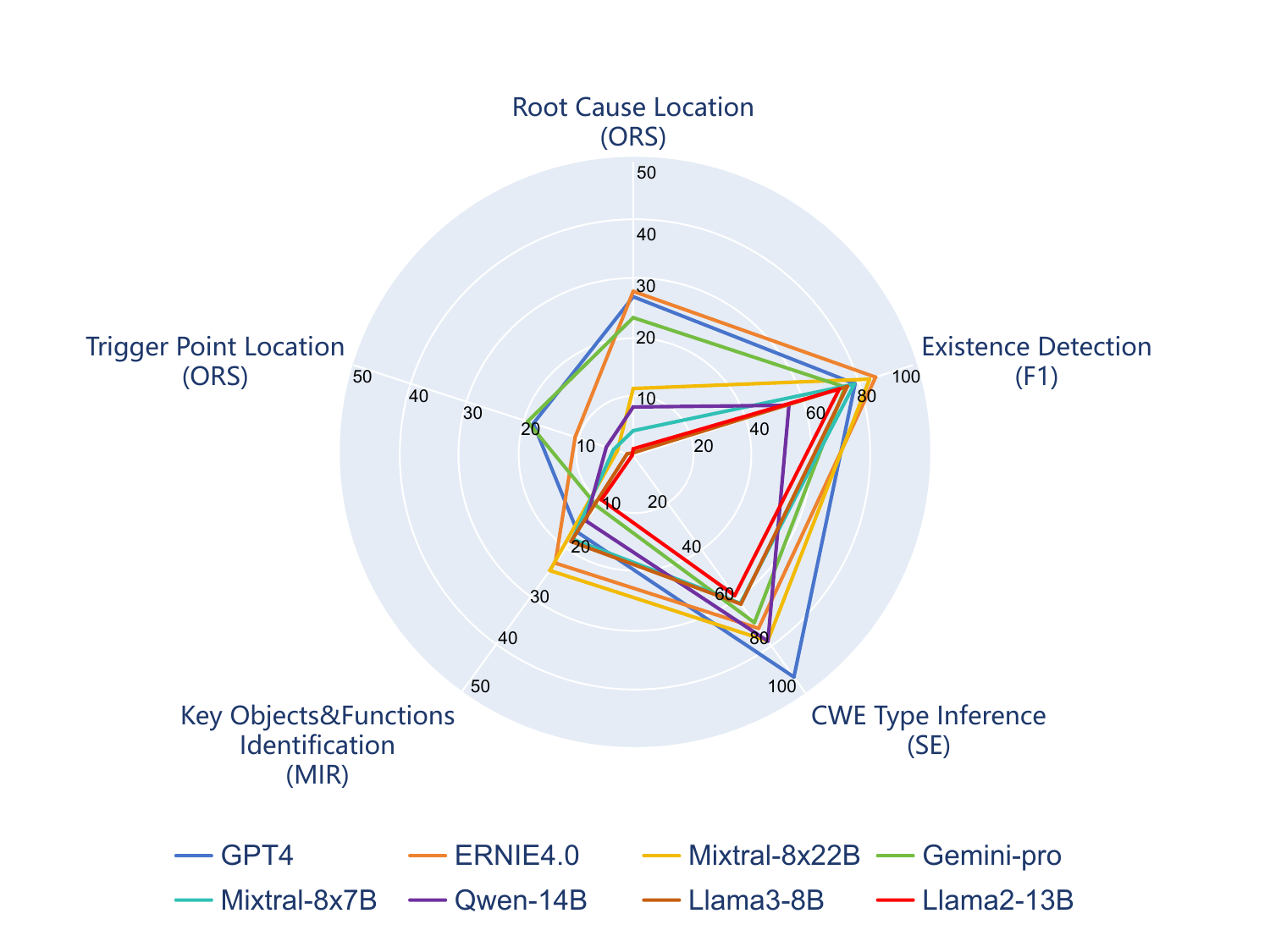}
    \caption{Top 8 LLMs' ability on Vulnerability Detections. Our benchmark consisting of five vulnerability analysis related tasks of increasing difficulty. The figure shows that existing LLMs perform well on simple analysis tasks such as vulnerability existence detection and CWE type inference, while on specific vulnerability related tasks, although performance varies from LLM to LLM, the overall performance is not yet satisfactory.}
    \label{fig:llm-ability}
\end{wrapfigure}

The task of vulnerability detection presents new challenges for LLMs. Firstly, programs with vulnerabilities in real-world scenarios tend to be large, and LLMs process these programs purely as text. Directly inputting such extensive programs into LLMs and expecting them to identify vulnerabilities is a daunting task. Secondly, even if a model can determine the presence of vulnerabilities within a program, we also need to know the types of vulnerabilities present. Different types of vulnerabilities have varying conditions for being triggered or exploited, which can provide security researchers with critical information for prioritizing patching efforts. Moreover, it's essential to ascertain whether the model can accurately locate the vulnerabilities, including the root causes and the specific locations where they are triggered. This capability significantly enhances the accuracy of vulnerability identification. These tasks become progressively more challenging, but the latter ones offer more valuable information for vulnerability analysis.

In this work, we construct a multi-task benchmark test centered around the specific concrete requirements of vulnerability analysis to evaluate the capability of the LLMs in vulnerability detection at multiple levels. We evaluate a total of 17 LLMs, including three closed-source models. The smallest model parameter with known parameters is 6B and the largest is 70B. As show in Figure \ref{fig:llm-ability}, we designed 5 tasks of progressive difficulty to progressively analyze the performance of the LLMs in existence detection, type classification, and vulnerability location. Our benchmark includes both artificially constructed datasets and real-world datasets. Furthermore, the programs in our benchmark dataset are written in C/C++, as a vast number of open-source libraries and kernels are developed in C/C++, and due to issues such as memory management, high-risk vulnerabilities like memory leaks frequently occur in C/C++ programs. Therefore, the capability of LLMs to detect vulnerabilities in large-scale C/C++ programs is highly representative.  


Our main contributions are threefold, \textbf{1): High-quality benchmark:}  We present a comprehensive benchmark meticulously crafted to evaluate the vulnerability detection abilities of large-scale models. The benchmark comprises a refined vulnerability dataset and incorporates five distinct assessment tasks, each focusing on different aspects of vulnerability analysis. \textbf{2): Comprehensive evaluations:} We conducted a comprehensive evaluation of the vulnerability detection capabilities of 17 existing large language models across 5 tasks. \textbf{3)Thorough analysis and new findings:} We performed an in-depth analysis of the evaluation results from various perspectives, shedding light on the strengths and limitations of existing large language models in vulnerability detection. Our findings lay the foundation for advancing the understanding and application of large language models in the domain of vulnerability detection.

\section{Related Work}
\subsection{Software Vulnerability and Detection Method.} Vulnerability is a flaw or weakness in a software program that can be exploited by an attacker to perform unauthorized actions within a computer system. Vulnerability detection involves identifying weaknesses in software programs that can be exploited by threats to compromise the security, functionality, or management of the software system. Researchers have proposed a large number of methods for identifying vulnerabilities in programs, which can be divided into two main categories: static analysis vulnerability detection\cite{beller2016analyzing} and dynamic analysis methods\cite{survey1}.

Static analysis\cite{lipp2022empirical}\cite{stefanovic2020static} typically relies on feature collection and templates to provide fast processing speeds. Different static analysis tools\cite{githubCodeQL}\cite{dwheelerFlawfinderHome}\cite{sourceforgeCppcheckTool} can deal with different types of vulnerabilities. However, due to the lack of contextual information from the execution environment, it often results in a high rate of false positives\cite{lipp2022empirical}, meaning that many vulnerability alerts are not true vulnerabilities, requiring manual verification of these alerts.


Dynamic analysis, such as fuzzing tests\cite{survey2, coredumpAmericanFuzzy}, explores as many parts of a program as possible by constructing a large diversity of program inputs, while dynamic fuzzing tests require the execution of a program to accurately identify vulnerabilities by directly observing the behavior of the vulnerabilities during execution. However, dynamic analysis is difficult and computationally expensive\cite{8835316} to explore all states of a large-scale program due to the complexity of the program, such as a large number of loops and branches. For security analysts, the operational costs are prohibitive. 

\subsection{Deep Learning in Vulnerability Detection} 
Deep learning methods\cite{li2018vuldeepecker} transform program code into code slices\cite{li2021sysevr}  or graphical structures\cite{mirsky2023vulchecker}, such as CFGs, DFGs, and PDGs, and train them using LSTM\cite{li2018vuldeepecker} or graph neural networks\cite{mirsky2023vulchecker}. These methods offer higher accuracy based on the learned features, significantly reducing the complexity involved in traditional dynamic and static analyses. However, these methods are often limited by the effect of the models themselves and the datasets used. Performance can significantly decline when the distribution of vulnerabilities in the datasets do not match real-world scenarios. Studies\cite{chakraborty2021deep} have shown that deep learning's performance in practical settings can dramatically decrease due to issues with the datasets.

Large language models are trained on extensive datasets that include program codes, vulnerability descriptions, and numerous code commits from open-source libraries, providing a certain level of understanding of various forms of vulnerabilities. These models treat program code as textual input, thereby rapidly assisting in vulnerability analysis. Despite these capabilities, research\cite{khare2023understanding} indicates that current large language models face limitations when dealing with complex programs in real-world scenarios. These models may not fully comprehend the complex logic and structure of advanced programs, necessitating further optimization and adjustment to adapt to real-world applications.

Recent studies also utilize LLMs for vulnerability detection, introducing new datasets\cite{chen2023diversevul}\cite{ding2024vulnerability} or benchmark\cite{gao2023far} focused on binary classification\cite{chen2023diversevul}\cite{khare2023understanding} or multi-classification \cite{gao2023far} of input program. However, these studies typically use function-level inputs, which do not accurately represent real-world vulnerability analysis scenarios, where vulnerabilities and their root causes, along with trigger points, often span multiple functions. Additionally, the sources of data used to construct benchmarks are crucial. Research indicates\cite{ding2024vulnerability}\cite{croft2023data} that the quality of existing datasets varies, influenced by the collection methods and their authenticity, whether synthetically generated or derived from real-world scenarios. The quality of labeling also varies across these datasets. These factors must be considered to ensure the reliability and validity of benchmarks.

The work  \cite{ullah2024llms}  proposes a framework for testing the capabilities of a given LLM as a security assistant across eight distinct dimensions. It includes the design of a multi-classification task (yes/no/n/a) for evaluating accuracy scores. Additionally, the study introduces more evaluation metrics related to the robustness of performing this task. However, it is important to note that the "root cause" identified in their work is not the specific program statement or expressions but rather the textual reasoning response generated by the LLMs. When a program contains hundreds of lines of code, such textual reasoning responses cannot precisely pinpoint the location of the vulnerability, which would be more valuable for security analysts.

Overall, the works \cite{fu2023chatgpt} \cite{purba2023software} \cite{zhou2024large} focus on evaluating large model vulnerability analysis capabilities. There are common problems at that time, firstly, the number of models issmall, and the evaluated models only focused on GPT4, GPT3.5 and a small number of other models such as CodeBert. Secondly, the metrics are relatively simple, and only evaluate whether the large model LLM could perform vulnerability detection on functions or code fragments. Third, there is no detailed discussion and design of the review dataset, e.g., long code, data labelling quality.

\begin{figure}[htb]
    \centering
    \includegraphics[width=1\linewidth]{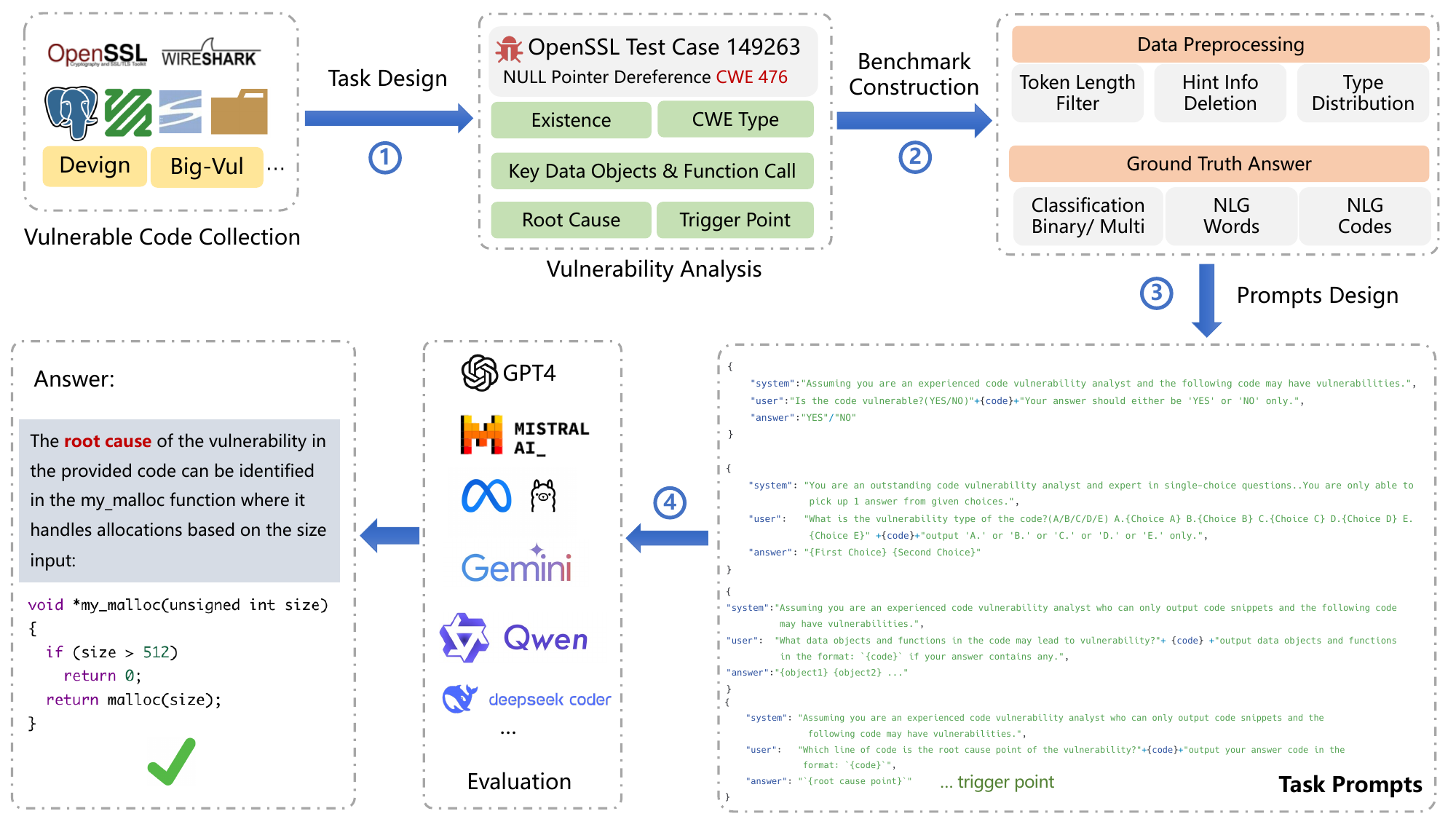}
    \caption{Overview of VulDetectBench Construction. We collect publicly available vulnerability databases to form the initial dataset and design five tasks with increasing difficulty to build the benchmark by analyzing the vulnerabilities. We ensure the quality of the benchmark through data processing, including the compatibility of the  context size limit, the deletion of obvious hints, and the distribution of the vulnerability types, and at the same time construct the ground truth answer. Next, we construct the prompts corresponding to the five tasks, input them to 17 LLMs for evaluation, and finally derive the analysis results. }
    \label{fig:pipelinefinal}
\end{figure}

\section{Benchmark Construction}

Figure \ref{fig:pipelinefinal} shows the overall process of VulDetectBench from construction to evaluation, and next we describe the dataset construction and specific task design respectively.

\paragraph{Data Source}
In this part, our goal is to construct a benchmark from high-quality datasets of code vulnerabilities. As shown in Table \ref{tab:data_source}. We collect real-world datasets from nine popular open source projects and a batch of datasets from the National Institute of Standards and Technology Software Assurance Reference Dataset (NIST SARD), which provides rich annotated data on real projects's vulnerabilities. Each entry in the dataset includes a detailed classification, description and key information about the code associated with the formation and triggering of the vulnerability, facilitating the design of various tasks. We also include two open source datasets: Big-Vul\cite{10.1145/3379597.3387501} and Devign\cite{zhou2019devign}.

In constructing the benchmark, we prioritize data quality, referencing the work \cite{croft2023data} to ensure the reliability of our review findings, which is not the focus in other datasets. The definition of "Accuracy" as used in\cite{croft2023data} and in Table \ref{tab:taskdesign} refers to the extent to which the data attributes accurately represent the true value of the intended attribute of a concept or event. Both the Big-Vul and Devign datasets rely on vulnerability-related information scraped from GitHub commits or links to vulnerability patches for data labeling. And, after a detailed investigation, the work\cite{croft2023data} find that the labels in these two datasets are not 100\% accurate. Moreover, the Big-Vul and Devign datasets only provide labels for the existence and CWE type of vulnerability, without offering details on the root cause or trigger points. Consequently, we only selecte parts of these datasets for constructing Task 1 and Task 2 in our benchmark. Given the presence of false positives in these datasets, we exclusively included data labeled as "no vulnerability." The remaining datasets used in our benchmark are sourced from real projects, where the root cause or trigger point has been clearly annotated, ensuring 100\% correct labeling. More detailed information about the datasets can be found in the \textbf{Appendix A.1}.

\begin{table}[htb]
	\centering
        \caption{Data Source. The Type in the table indicates whether the source of the dataset is synthetic or real data, while the related work\cite{croft2023data} points out that the accuracy of vulnerability labelling in the dataset varies, and overall it is a false alarm, so we selecte the data labelled as no vulnerability as part of the data for our benchmark in Devign and Big-Vul.}
	\resizebox{\linewidth}{!}{
		\begin{tabular}{l|c|c|c|c|c}
			\hline
			Dataset Name & Version & Project &Type& Entries & Accuracy\\ \hline
			Apache Subversion \cite{nistapachesubversion} & 1.8.3 & Subversion & Real & 638 &1.000  \\
			Wireshark \cite{nistwireshark} & 1.10.2 & WireShark & Real & 637  &1.000\\
			Tree \cite{tree1.7.0} & 1.7.0 & Tree  & Real & 380  &1.000\\
			PostgreSQL \cite{nistpostgresql924} & 9.2.4 & PostgreSQL  & Real & 637  &1.000\\
			OpenSSL\cite{openssl} & 1.0.1e & OpenSSL  & Real & 636  &1.000\\
			GNU Grep\cite{grep214} & 2.14 & Grep  & Real & 380  &1.000\\
			FFmpeg \cite{ffmepg22} & 1.2.2 & FFmpeg  & Real & 637  &1.000\\
			Gimp\cite{gimp288} & 2.8.8 & GIMP & Real & 637  &1.000\\
			\text{Juliet C/C++} \cite{julietcc} &  \makecell{1.3.1 \\ with extra support}  & SARD & Synthetic & 64099  &1.000\\
			Devign \cite{zhou2019devign} & - & FFmpeg, Qemu, Wireshark  & Real & 27318  &0.800\\
			Big-Vul \cite{10.1145/3379597.3387501} & - & Multiple sources  & Real & 43523  &0.543\\ \hline
		\end{tabular}
	}

	\label{tab:data_source}
\end{table}
\paragraph{Task Design}
VulDetectBench is a comprehensive benchmark tailored to large language models for detecting vulnerabilities in code. The benchmark consists of five tasks of increasing difficulty that provide an in-depth assessment of the vulnerability detection capabilities of LLMs: Task 1 involves a binary classification to determine the presence of code vulnerabilities. Task 2 is a multi-class classification task aimed at identifying the CWE (Common Weakness Enumeration) classification of code vulnerabilities. Starting from Task 3, we increased the difficulty level to explore whether models can truly understand the specific content of vulnerabilities. Tasks 4 and 5 further assess the model's ability to identify the root causes and trigger points of vulnerabilities, respectively. These tasks are critical for in-depth vulnerability analysis and represent the most challenging aspects, testing the models' semantic understanding of program vulnerabilities comprehensively.


\begin{table*}[h!]
		\centering
            \caption{Task Design. VulDetectBench contains five tasks, and the number of entries for evaluating LLMs varies in different tasks. We control the length of the evaluation data token to be under 4K. The min tokens are smaller because Task 1 and Task 2 contain manually constructed datasets. The data in Tasks 3 to 5 are constructed with the vulnerability dataset of real engineering, which contains the same 100 data entries, and also the 100 data entries are fully covered in Tasks 1 and 2.}
		\resizebox{\linewidth}{!}{
			\begin{tabular}{llcccc}
				\hline
				No. & Task  & \makecell{Number \\ of Entries}  & \makecell{CWE \\  Types}   & \makecell{TOP 5 \\ CWE Type}  & \makecell{ Min - MAX \\ tokens} \\ 
				\hline
				1 & Vulnerability Existence Detection  & 1000 & 48 & 78 | 90 | 23 | 36 | 114 & 50 - 3493  \\
				2 & Vulnerability CWE Type Inference  & 500 & 48 & Same Above & 265 - 3372  \\
				3 & Key Objects and Functions Identification  & 100 & 38 & 476 | 191 | 88 | 78 | 89 & 1017 - 3269  \\
				4 & Vulnerability Root Cause Location  & 100 & 38 & Same Above & 1010 - 3262  \\
				5 & Vulnerability Trigger Point Location & 100 & 38 & Same Above & 1011 - 3363  \\ \hline
			\end{tabular}
		}

		\label{tab:taskdesign}
	\end{table*}

As shown in Table \ref{tab:taskdesign}. To enhance the precision of assessing the vulnerability analysis capabilities of Large Language Models (LLMs), the benchmarks for Tasks 1 and 2 are designed to contain identical datasets, each comprising 1,000 data entries. Similarly, Tasks 3, 4, and 5 are standardized with exactly the same sample data, encompassing 100 data entries for each task. This uniformity allows for direct comparison across tasks and facilitates tracing specific samples to determine the maximum difficulty level the models can handle. 

Using our benchmark, we feed the full code into LLMs. For task 1 and task 2, the  Min-max tokens of data are 45-4051 and 132-4121, because we sample some data from Devign and Big-vul which contains many short functions. For task3, task4 and task5, all the data from real projects like Wireshark, Apache, and Openssl. The data in these datasets are very long, ranging from 1050-3300 as shown in Table\ref{tab:taskdesign}. 


\paragraph{Task 1: Vulnerability Existence Detection}
For Task 1, we examine the capability of existing LLMs to perform the relatively simple binary classification task of determining the presence of vulnerabilities in a program.  Vulnerable samples for this task are constructed from both real-world and synthetically generated datasets that are well-annotated and provide comprehensive relevant information. The test cases in task1 require the LLMs to output a "YES" or "NO" to indicate the existence of a vulnerability. During evaluation, the accuracy and F1 score of the models are assessed based on whether their output matches the correct answer.



\paragraph{Task 2: CWE Type Inference}
Specifically, for this task, we present five different options for each sample: the optimal answer, which is the sample's actual CWE type, and a suboptimal answer derived from the CWE-1000 VIEW hierarchy\cite{mitreCWE1000Research}, representing an ancestor node within four layers. Additionally, three incorrect choices are provided: two unrelated CWE types and one "No Vulnerability" option. Models must correctly identify the CWE type or select the correct option letter, with scoring based on two systems. The \textbf{Moderate Evaluation (ME)} awards 1 point for the optimal choice and 1 point for choosing the suboptimal without the optimal, whereas the \textbf{Strict Evaluation (SE)} grants 1 point for the optimal and 0.5 points for the suboptimal, with no points for other selections. This method scrutinizes the models' depth of understanding and accuracy in inference vulnerability types within complex C/C++ program structures.


\paragraph{Task 3: Key Data Objects and Functions Identification} Identifying key data objects and function calls associated with vulnerabilities is crucial in vulnerability analysis. For instance, analyzing CWE 476 buffer overflow vulnerabilities involves examining the buffer and related memory read-write functions. LLMs' ability to identify these key elements is essential for accurate vulnerability analysis. In Task 3, both Macro Recall (MAR) and Micro Recall (MIR) metrics are used. MIR is particularly effective in mitigating fluctuations caused by sparse labels in the ground truth, providing a more stable evaluation of model performance.
\begin{align}
	\text{MAR} =\frac{1}{n}\sum_{i=1}^n(\frac{TP_i}{TP_i+FN_i}) \quad\quad \text{MIR} =\frac{\sum_{i=1}^n(TP_i)}{\sum_{i=1}^n(TP_i+FP_i)}
\end{align}

\paragraph{Task 4: Root Cause Location}
For Task 4 in our benchmark, we focus on evaluating LLMs' ability to locate the root cause of a vulnerability within a program's codebase, a crucial aspect for precise vulnerability analysis. We use a data set where each vulnerability's root cause is uniquely labeled, and employ a prompt designed to compel the model to identify the specific code region associated with the root cause. By extracting these identified regions, we calculate the recall of the model's responses, thus measuring how effectively it can pinpoint the root causes within the textual code data.

\paragraph{Task 5 : Trigger Point Location}
In Task 5, we examine the LLMs' capacity to identify the precise trigger point of a vulnerability, typically localized to a few lines or a single line of code. This task challenges the models to comprehend extensive code text and accurately determine the specific trigger point, which is essential for effective vulnerability localization. The task involves a similar setup to Task 4, with a prompt that requires the model to specify the code region during its response. The performance is evaluated by calculating the recall of the model's output in identifying the exact lines or line where the vulnerability is triggered.
For both Task 4 and Task 5, the primary metric for evaluation is line-of-code recall. This involves segmenting identified code snippets from the model's output, analyzing them by lines, and assessing the recall of the lines that correspond to the standard answers, using a macro-mean recall for the final metric. For Task 5, trigger point code lines are extracted and analyzed similarly. Additionally, an alternative recall metric is computed—defined by the intersection over union of the model’s output and the true labels—to ensure a balanced assessment of the model’s accuracy and precision in both identifying and localizing vulnerabilities.
\begin{align}
	\text{URS} =\frac{1}{n}\sum_{i=1}^n(\frac{IL_i}{ROL_i}) \quad\quad \text{ORS} =\frac{1}{n}\sum_{i=1}^n(\frac{IL_i}{UL_i})
\end{align}
URS stands for Union Recall Score and IL stands for the intersection lines of code between  LLMs' output and the ground truth answers. ROL stands for Result Output Lines and UL stands for Union Lines between LLMs output and ground truth answers. ORS is designed to detect whether the model complies well with the instructions and only outputs codes, and on the other hand, to mitigate the large number of false alarms introduced by the model when it outputs a large number of codes to increase its score, which is difficult to judge.

\section{Experiment}

\subsection{Setup}


We selected 17 models from 10 different families in the experiment, including proprietary models like GPT-4, Google Gemini-Pro, and Ernie 4.0. The models varied in size, with parameters ranging from 6B to 70B. Mixtral 8 * 22B, an open-source model, had the highest parameter count. We ensured that at least two models with different parameter sizes are chosen from each family for comparison. Notably, Deepseek and CodeLlama underwent code-specific pre-training.

The computational resources are standardized using NVIDIA GeForce RTX 4090 graphics cards. Single RTX 4090 units are used for models up to 7B, while two RTX 4090s are used in parallel for larger models (13B-14B) to maintain optimal conditions and avoid performance bottlenecks.

\subsection{Overall Results}
As presented in Table \ref{tab:overall_result}, we evaluate the performance of 17 Large Language Models (LLMs) across five tasks, using two distinct metrics for each task. For Task 1, we measure Accuracy and F1 Score. In this task, Ernie 4.0, a proprietary model, outperforms all others, achieving the highest Accuracy of 85.01\% and F1 Score of \textbf{86.65\%}. Among the open-source models, Mixtral-8*22B shows the best performance in Accuracy. For models with a smaller number of parameters, Meta-Llama-8B-instrcut demonstrates superior performance, achieving an F1 Score of \textbf{76.53\%}.
\begin{table}[htb]
	\centering
	\caption{Overall Results. For each of the five tasks in the VulDetectBench, we evaluate 17 LLMs, three of which were closed-source models, using two metrics. The specific types of these models are described in detail in the Appendix.}
	\resizebox{\linewidth}{!}{
		\begin{tabular}{lcccccc}
			\toprule
			Model  &  Size  & \makecell{Task1 \\ ACC   |    F1 } &  \makecell{Task2  \\   SE | ME} &   \makecell{Task 3   \\ MAR|   MIR }&  \makecell{Task 4  \\  URS  | ORS  } & \makecell{Task 5  \\ URS | ORS   } \\
			\midrule
			GPT4 & -                & 71.43  |    79.30          & \textbf{92.96 |   95.17}                &  20.21 |   16.07 & \textbf{24.26 |   27.07} & \textbf{13.00 |   17.85} \\
			ERNIE4.0    & -                & \textbf{85.01 |   86.65} & 72.50 |   70.00                         & 27.87 |   22.54 & \textbf{11.77 |   27.99} & \textbf{22.43 |   10.38} \\
			Gemini-pro  & -                      & 73.10 |   75.74          & 70.10 |   77.00                         & 13.03 |   10.55 & 14.64 |   23.51          & 07.56 |   18.89          \\
			Deepseek & 7B & 71.30 |   61.16          & 37.60 |   40.20                         & 17.81 |   13.67 & 05.36 |   09.22          & 04.30 |   08.83          \\
			Qwen & 7B                 & 62.30 |   43.31          & 60.10 |   62.20                         & 13.63 |   11.15 & 04.34 |   10.09          & 05.56 |   10.95          \\
			Qwen & 14B              & 69.10 |   55.67          & 77.90 |   82.80                         & 16.49 |   13.67 & 06.40 |   08.32          & 03.14 |   04.81          \\
			ChatGLM3 & 6B                    & 69.90 |   65.37          & 39.90 |   46.60                         & 00.16 |   00.12 & 00.19 |   01.33          & 00.42 |   01.12          \\
			Vicuna & 13B                & 57.60 |   31.17          & 67.30 |   74.40                         & 07.30 |   06.35 & 0 |   0                 & 0 |   0                \\
			Vicuna & 7B                & 48.60 |   65.27          & 31.30 |   42.00                          & 07.79 |   06.24 &  0 |   0                & 0 |   0                \\
			CodeLlama & 13B      & 47.90 |   58.81          & 32.80 |   44.60                         & 10.34 |   08.51 & 03.30 |   03.47          & 01.29 |   01.89          \\
			CodeLlama & 7B        & 36.40 |   53.37          & 36.70 |   41.60                         & 06.28 |   05.04& 01.69 |   02.55          & 00.69 |   01.31          \\
			Llama3 & 8B        & 69.40 |   76.53          & 62.30 |   68.40                         & 22.83 |   17.99 & 00.19 |   01.10           & 00.17 |   00.53          \\
			Llama2 & 7B              & 47.90 |   64.19          & 41.40 |   54.40                         & 11.47 |   09.59 & 0 |   0                 & 0 |   0                \\
			Llama2 & 13B           & 70.37 |   73.67          & 58.70 |   67.40                         & 10.99 |   09.23 & 00.50 |   01.20           & 00.10 |   00.20          \\
			Llama3 & 70B                     & 47.45 |   60.33          & 26.00 |   17.00                         & 09.18 |   07.43 & 0 |   0                 & 0 |   0                 \\
			Mixtral & 8*7B                  & 76.42 |   79.00          & 62.00 |   58.40                         & 21.61 |   17.51 & 01.66 |   04.28          & 05.74 |   03.51          \\
			Mixtral & 8 *22B                 & 81.82 |   84.47          & 77.80 |   74.80                       &  \textbf{30.26  |   24.10} & 17.49 |   11.46          & 06.17 |   02.83     \\    
			\bottomrule
		\end{tabular}
		\label{tab:overall_result}
	}
	
\end{table}

In Task 2, the performance of different LLMs varies significantly. Some models, such as GPT-4, Qwen-14B-Chat, and Vicuna-13B-v1.5, perform better than in Task 1, while the majority of the models exhibit a substantial decline in performance. It is important to note that, generally, using the Moderate metric leads to an improvement of approximately \textbf{5\%}, and in some cases, such as with Vicuna-7B-v1.5, the improvement reaches \textbf{12\%}. This suggests that the models' ability to predict vulnerability types is influenced by their training data, resulting in varied performance under different precision requirements of the same task.

In Task 3, the open-source model Mixtral-8*22B demonstrates a significant advantage, achieving a Macro Recall of 30.26\%. This performance level provides valuable insights for vulnerability analysis. Among the smaller models, Meta-Llama-8B-instruct also performs well, outperforming other open-source models by an average of more than 5\% in Macro Recall.

For Tasks 4 and 5, we employ two different metrics for evaluation, as outlined in Equations 2. We observe a significant performance decline across all models in these tasks, yet the disparity in effectiveness between models remains substantial. Some open-source models, such as the Vicuna, Llama families, and ChatGLM3-6B, struggle considerably with these tasks, nearly failing to accomplish them. GPT-4 and Ernie 4.0 exhibit the best performance in both tasks. However, we note that Ernie 4.0 experiences a decline of over \textbf{10\%} in LOC Recall based on intersections, and upon analyzing the actual output, we find that Ernie 4.0's responses contain a substantial amount of irrelevant content, which adversely affects practical vulnerability analysis. Based on this comprehensive analysis, GPT-4 emerges as the most effective model in these tasks.

Overall, we observe that existing LLMs excel in the binary classification task of detecting vulnerabilities in programs (Task 1), achieving best accuracy above 85\% for inputs up to 4K tokens. This performance is consistent across both open-source and proprietary commercial models. However, their performance declines in Task 2, which involves identifying specific CWE vulnerability types. Furthermore, all models significantly underperform in tasks that require identifying key data objects and function calls associated with vulnerabilities, as well as in accurately pinpointing the root causes and trigger points of these vulnerabilities. We also fintune the LLM to check if there are any boost of performance in \textbf{Appendix A.6}
\begin{figure}
	\centering
	\includegraphics[width=0.9\textwidth]{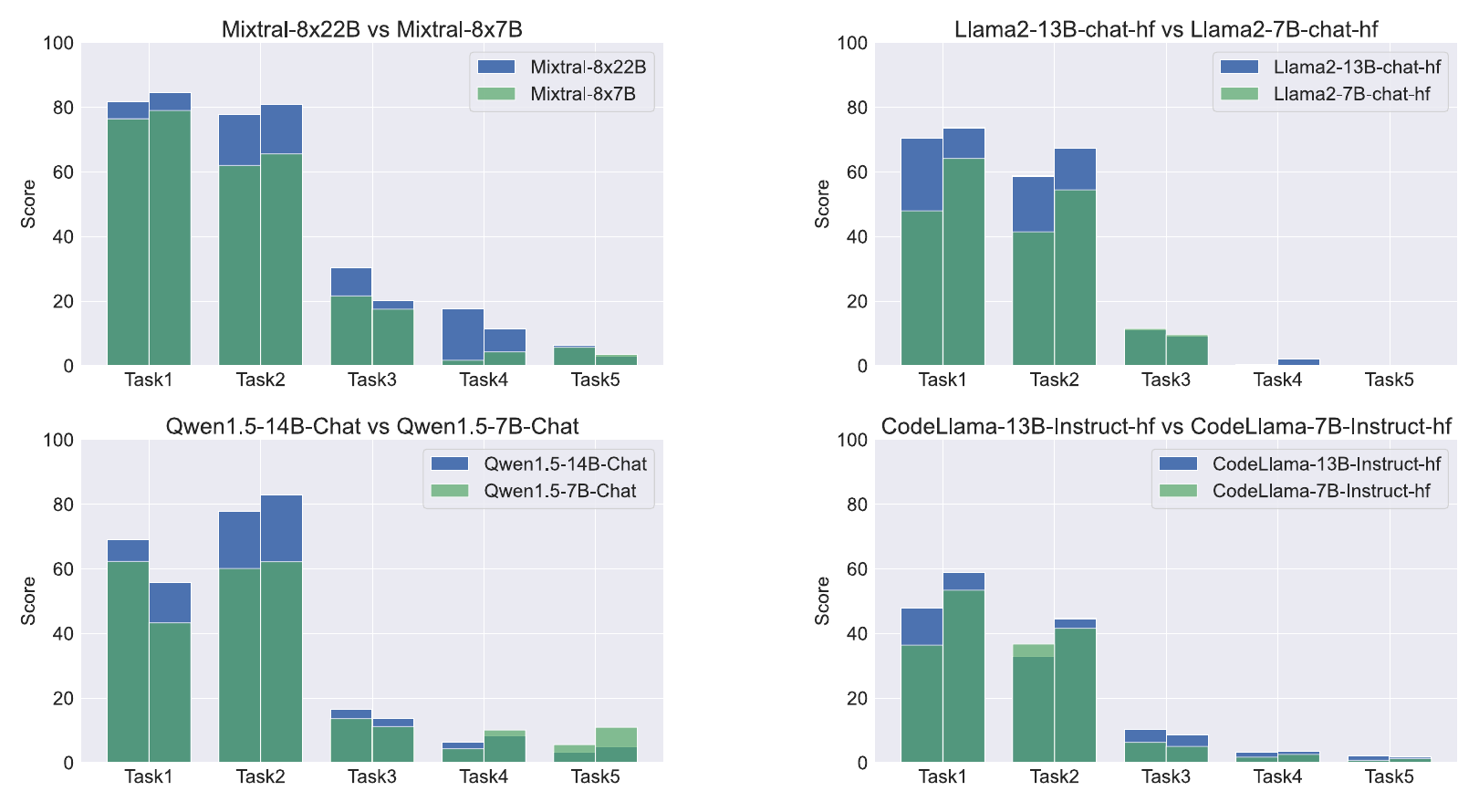}
	\caption{Performance comparison of different sizes within the same LLM family}
	\label{fig:same_model_battle}
\end{figure}	
\subsection{In-depth Analysis}
\paragraph{Model Size Influence}
In this section, we conduct a detailed analysis of models from the same family but of varying sizes on their performance across the aforementioned five tasks. Generally, it is observed that models with a larger number of parameters tend to perform better on these tasks. Specifically, we exemplify this trend using the models Mistral, Llama2, Qwen, and CodeLlama. As shown in Figure \ref{fig:same_model_battle}, for five LLM families, we analyze the impact of model parameter size on performance across different tasks. It is evident that models with larger parameters generally perform better across all tasks. In simpler tasks, the increase in parameter size correlates with greater performance improvements, with a maximum increase of 46.9\% in Task 1 and up to 115.0\% in Task 2. However, the extent of performance improvement varies under different metrics. For Task 1, the metric based on the F1 score is more stringent than the Accuracy metric, resulting in a lower percentage of improvement for the models, especially for the Vicuna-13b-v1.5, where the F1 Score decreased by 52.2\% compared to the same family's 7B model.
\paragraph{Semantic Understanding of Vulnerability}
To better understand the extent of the LLMs' understanding of vulnerabilities, we use 100 identical test cases from tasks one to five for analysis. Despite their ability to accurately determine the presence of vulnerabilities and even identify CWE types, this does not mean that the models truly understand the specific details of the vulnerabilities. For example, they are unable to accurately identify data objects and critical function calls that are closely associated with vulnerabilities, nor can they accurately locate root causes and trigger points. This suggests that large models assess the presence of vulnerabilities primarily from the perspective of high-level language features, and lack a deeper understanding of the specific mechanisms of vulnerability occurrence.


\begin{figure}[htbp]
    \begin{minipage}[t]{0.5\linewidth}
        \centering
        \subfigure[]{\label{fig:subfig1}\includegraphics[width=0.95\textwidth]{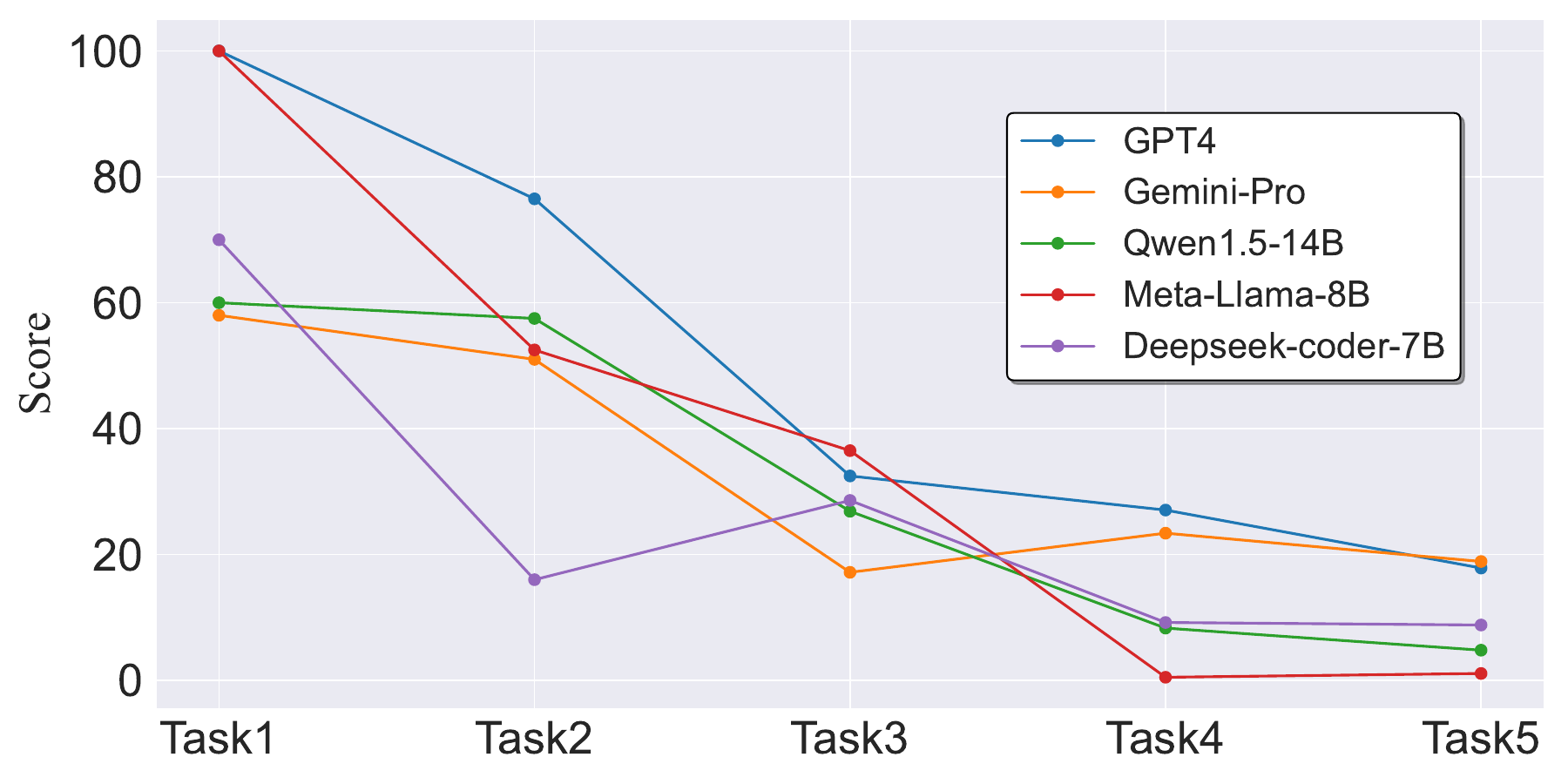}}
    \end{minipage}%
    \begin{minipage}[t]{0.5\linewidth}
        \centering
        \subfigure[]{\label{fig:subfig2}\includegraphics[width=0.95\textwidth]{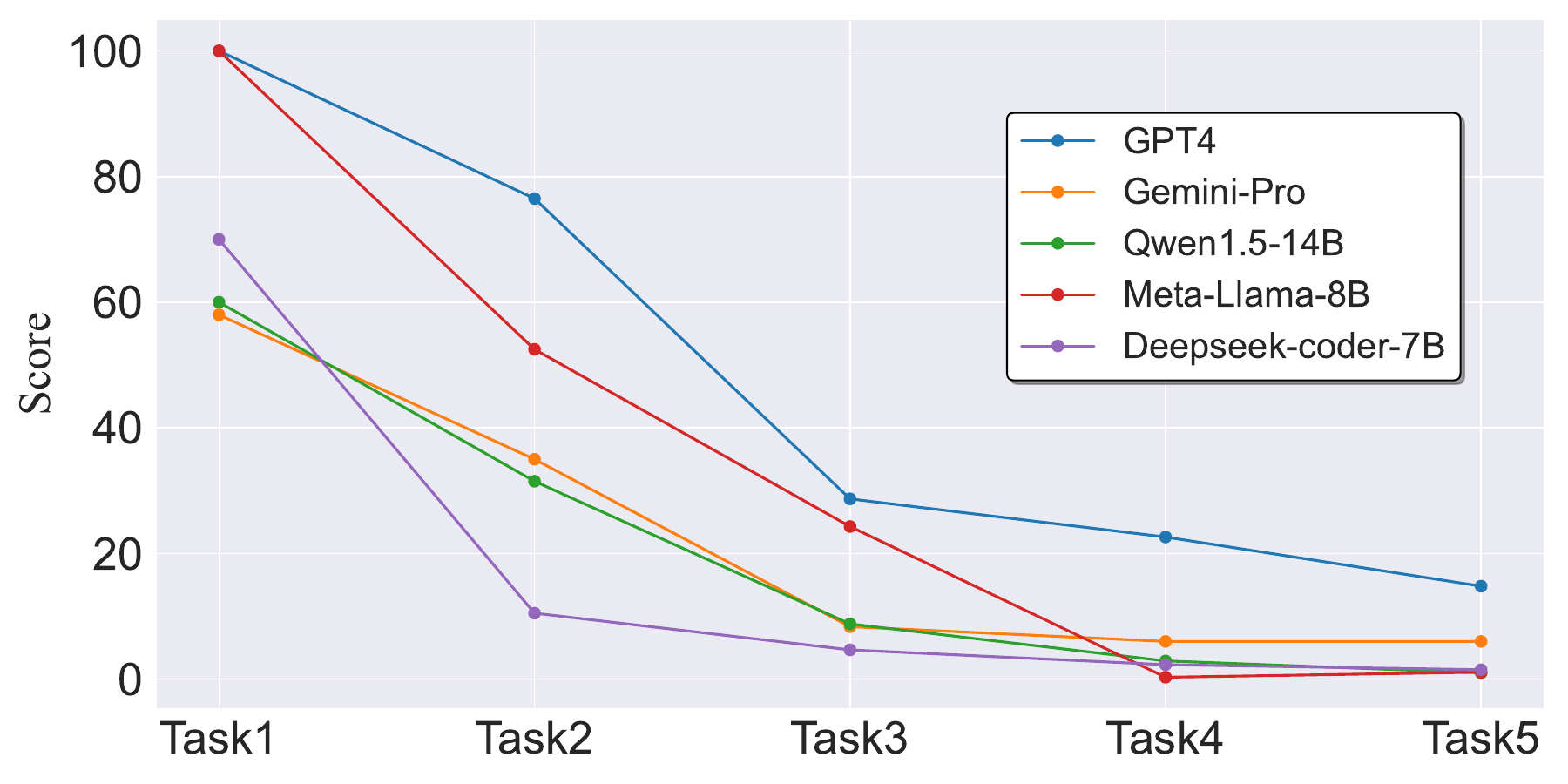}}
    \end{minipage}
    \caption{Performance comparison. \ref{fig:subfig1} represents the performance on 100 samples of different tasks. \ref{fig:subfig2} represents the performance of the model given that the previous task was performed correctly.(The quantity in Task 2 is the number that Task 1 gets right.)}
    \label{fig:compare}
\end{figure}


As shown in Figure \ref{fig:compare}. Through a horizontal comparison among different tasks assigned to models, it is evident that performance on simpler tasks does not fully reflect a model's capabilities in detecting code vulnerabilities. Taking GPT4 as an example, its performance in Task 1 suggests it can effectively identify the presence of vulnerabilities in code. However, its reduced performance in Task 2 indicates a biased judgment in detecting code vulnerabilities during Task 1. Further analysis of its performance in Tasks 3, 4, and 5 reveals that it struggles to accurately locate the trigger points, root causes, and critical data types associated with code vulnerabilities. Therefore, we conclude that the current capabilities of LLMs in vulnerability detection are limited and should be evaluated based on tasks of varying difficulty to accurately assess their ability to detect code vulnerabilities.

\paragraph{Performance on specific type of vulnerability}

We conduct the experiment by testing the scores for different model models against different CWE types of vulnerabilities in task4 and task5 (in this experimental setup, as long as some lines of the answer intersected with the ground truth, the answer was judged to be primed to locate the vulnerability, which was scored as a 1, and the rest were scored as a 0, and the percentage was calculated). As shown in Table \ref{tab:cwe_performance}.
\begin{table}[H]
    \centering
    \caption{Performance comparison of different models on Task 4 (Root Cause Location) and Task 5 (Trigger Point Location) for specific CWE types. The values represent the Unified Recall Score (URS) for each task, formatted as Task4|Task5. }
    \resizebox{\linewidth}{!}{
        \begin{tabular}{lccccc}
            \toprule
            CWE for Model (TASK4|TASK5) & GPT4 & Gemini-pro & Deepseek & ChatGLM3 & QWEN(7B) \\
            \midrule
            CWE-78 OS Command Injection & 0.83|0.83 & 0.17|0.33 & 0.33|0.17 & 0.00|0.00 & 0.00|1.00 \\
            CWE-88 Argument Injection & 0.71|0.14 & 0.71|0.14 & 0.29|0.57 & 0.00|0.29 & 0.57|0.57 \\
            CWE-191 Integer Underflow & 0.00|0.14 & 0.00|0.57 & 0.00|0.57 & 0.00|0.14 & 0.29|0.86 \\
            CWE-89 SQL Injection & 0.83|0.00 & 0.33|0.33 & 0.00|0.33 & 0.00|0.17 & 0.00|0.50 \\
            CWE-476 NULL Pointer Dereference & 0.20|0.25 & 0.40|0.10 & 0.30|0.10 & 0.05|0.00 & 0.60|0.20 \\
            CWE-775 Missing Release of File Descriptor & 0.60|0.80 & 1.00|0.20 & 1.00|0.20 & 0.00|0.00 & 0.80|0.60 \\
            CWE-682 Incorrect Calculation & 0.00|0.00 & 0.25|0.50 & 0.25|0.00 & 0.00|0.00 & 0.00|0.25 \\
            CWE-120 Classic Buffer Overflow & 1.00|0.75 & 0.50|1.00 & 0.00|0.25 & 0.25|0.25 & 0.00|0.75 \\
            CWE-190 Integer Overflow or Wraparound & 0.00|0.00 & 0.33|0.00 & 0.00|0.33 & 0.00|0.00 & 0.00|0.68 \\
            \bottomrule
        \end{tabular}
    }
    \label{tab:cwe_performance}
\end{table}
We also conduct the following experiment: for each CWE type in TASK2 (top 10 in number of 500 test cases, num is the specific number) the model corresponds to the proportion of correctly classified, as shown in Table \ref{tab:cwe_classification_performance}.
\begin{table}[H]
    \centering
    \caption{Performance comparison of different models on Task 2 (Vulnerability CWE Type Inference) for the top 10 most frequent CWE types. The table shows the proportion of correctly classified instances for each CWE type across various models. The 'Num' column indicates the number of test cases for each CWE type out of 500 total test cases.}
    \resizebox{\linewidth}{!}{
        \begin{tabular}{lccccccc}
            \toprule
            CWE TYPE & Num & GPT4 & Gemini-pro & Deepseek & ChatGLM3 & Qwen(7B) \\
            \midrule
            CWE-78 OS Command Injection & 75 & 1.00 & 0.77 & 0.66 & 0.24 & 0.77 \\
            CWE-90 LDAP Injection & 74 & 0.99 & 0.78 & 0.63 & 0.14 & 0.79 \\
            CWE-23 Relative Path Traversal & 72 & 1.00 & 0.77 & 0.42 & 0.09 & 0.80 \\
            CWE-36 Absolute Path Traversal & 68 & 1.00 & 0.91 & 0.26 & 0.13 & 0.88 \\
            CWE-114 Process Control & 43 & 1.00 & 0.48 & 0.28 & 0.22 & 0.16 \\
            CWE-15 External Control of System or Configuration Setting & 37 & 1.00 & 0.97 & 0.43 & 0.07 & 0.70 \\
            CWE-476 NULL Pointer Dereference & 26 & 0.62 & 0.44 & 0.14 & 0.04 & 0.33 \\
            CWE-89 SQL Injection & 8 & 1.00 & 0.88 & 0.13 & 0.00 & 0.25 \\
            CWE-191 Integer Underflow (Wrap or Wraparound) & 8 & 0.50 & 0.50 & 0.00 & 0.00 & 0.31 \\
            CWE-88 Argument Injection & 7 & 0.93 & 0.79 & 0.14 & 0.14 & 0.43 \\
            \bottomrule
        \end{tabular}
    }
    \label{tab:cwe_classification_performance}
\end{table}
The results of these two experiments indicate that LLMs exhibit varying abilities in analyzing different specific types of vulnerabilities. In Task 2, the closed-source model demonstrates superior overall performance, though the open-source model is also capable of handling certain tasks, particularly for vulnerabilities with more distinctive features, such as CWE-78 and CWE-23. LLMs show competence in identifying CWE types in these cases. However, for Tasks 4 and 5, only GPT-4 is able to provide relatively valuable vulnerability localization results for specific tasks. For vulnerability types like CWE-191, which require procedural analysis and possess more hidden characteristics, LLMs are not yet capable of delivering valuable information.

\paragraph{Comparison with Traditional Tools}
Vulnerability analysis of programs using LLM has the intuitive advantage of being less dependent on the integrity of the program. Traditional static program analysis tools such as Infer require compilation of the program, which transforms program syntax trees (ASTs), program control flow graphs (CFGs), and other forms of program representation. CodeQL\cite{githubCodeQL} requires the use of manually constructed databases to perform vulnerability matching. This does not allow for rapid vulnerability analysis, and we choose a commonly used tool  flawfinder\cite{dwheelerFlawfinderHome} that can directly analyze portions of a program's code for comparative experiments. This tool can complete five tasks in VulDetectBench, but none of them give correct results and produce a large number of false positives. More detailed descriptions are provided in the \textbf{Appendix A.4}.



\section{Conclusion, Limitations and Societal Impacts}
\textbf{Conclusions} Our study introduces a new benchmark for evaluating LLMs on a range of tasks using high-quality real data and long program files as input. By focusing on tasks of increasing complexity, we have gained valuable insights into the capability boundaries of both open-source and proprietary LLMs. While these models perform well on basic vulnerability detection tasks—effectively detecting the presence of vulnerabilities and inferring CWE types—they fall short in more specialized tasks, such as identifying key data objects and accurately determining the root cause of vulnerabilities. Our research emphasizes the need for precision in root cause identification, particularly when dealing with extensive codebases, where pinpointing exact vulnerability locations is essential. Additionally, our evaluation methodology directly assesses the LLMs' ability to locate root causes, trigger points, or key data objects, offering a more granular and precise assessment of their performance. Although LLMs can serve as useful auxiliary tools for vulnerability analysis, their current capabilities are insufficient to replace traditional tools or provide comprehensive analytical support in more complex scenarios.

\textbf{Limitations} While our benchmark provides a comprehensive evaluation of LLMs on vulnerability detection tasks, there are opportunities for future work to expand the dataset to include more programming languages and a wider range of vulnerability types.

\textbf{Societal impacts} Our benchmark can facilitate the development of large language models for vulnerability detection. Meanwhile, Using large language models for vulnerability detection may lead to malicious exploitation and over-reliance, necessitating responsible development and oversight.


\bibliographystyle{unsrt}
\bibliography{refbib}

\newpage
\appendix

\section{Appendix}

\subsection{Data Source Description}

\textbf{Juliet C/C++:} A dataset of 64,099 C/C++ code samples showcasing vulnerabilities and their fixes. Vulnerabilities are marked with comments like \texttt{POTENTIAL FLAW:XXX} and can be toggled via macro definitions. Each sample includes explanations of Good Source, Bad Source, and Bad Sink.

\textbf{OpenSSL:} Contains 636 repair patches with vulnerabilities annotated using comments like \texttt{STONESOUP: TRIGGER-POINT (...)}, \texttt{ STONESOUP: CROSSOVER-POINT (...)} and \texttt{STONESOUP: AFTER-POINT (...)}in the codes. Each sample has an associated \texttt{manifest.sarif} file detailing vulnerability descriptions and CWE categories.

\textbf{BigVul:} Features 3,754 CVE-related samples from 2002 to 2019, each detailed with 21 features. The dataset provides comparisons of code before and after fixes, vulnerability classifications, and additional metadata.

\textbf{Devign:} Utilizes data from Linux, FFmpeg, Qemu, and Wireshark to predict the existence of vulnerabilities. Each project offers different functionalities and common vulnerabilities, such as memory corruption in Linux, DoS and code execution in Qemu, memory leaks in Wireshark, and overflows in FFmpeg.

The following datasets contain repair patches with detailed manual annotations, using the same annotation method and vulnerability information format as the OpenSSL dataset:

\begin{itemize}
    \item \textbf{Apache Subversion:} Contains 638 repair patches.
    \item \textbf{GNU Grep:} Contains 380 repair patches.
    \item \textbf{Wireshark:} Contains 637 repair patches.
    \item \textbf{Tree:} Contains 637 repair patches.
    \item \textbf{PostgreSQL:} Contains 637 repair patches.
    \item \textbf{Gimp:} Includes 637 repair patches.
    \item \textbf{FFmpeg:} Includes 637 repair patches.
\end{itemize}

\begin{figure}[ht]
    \centering
    \includegraphics[width=0.8\linewidth]{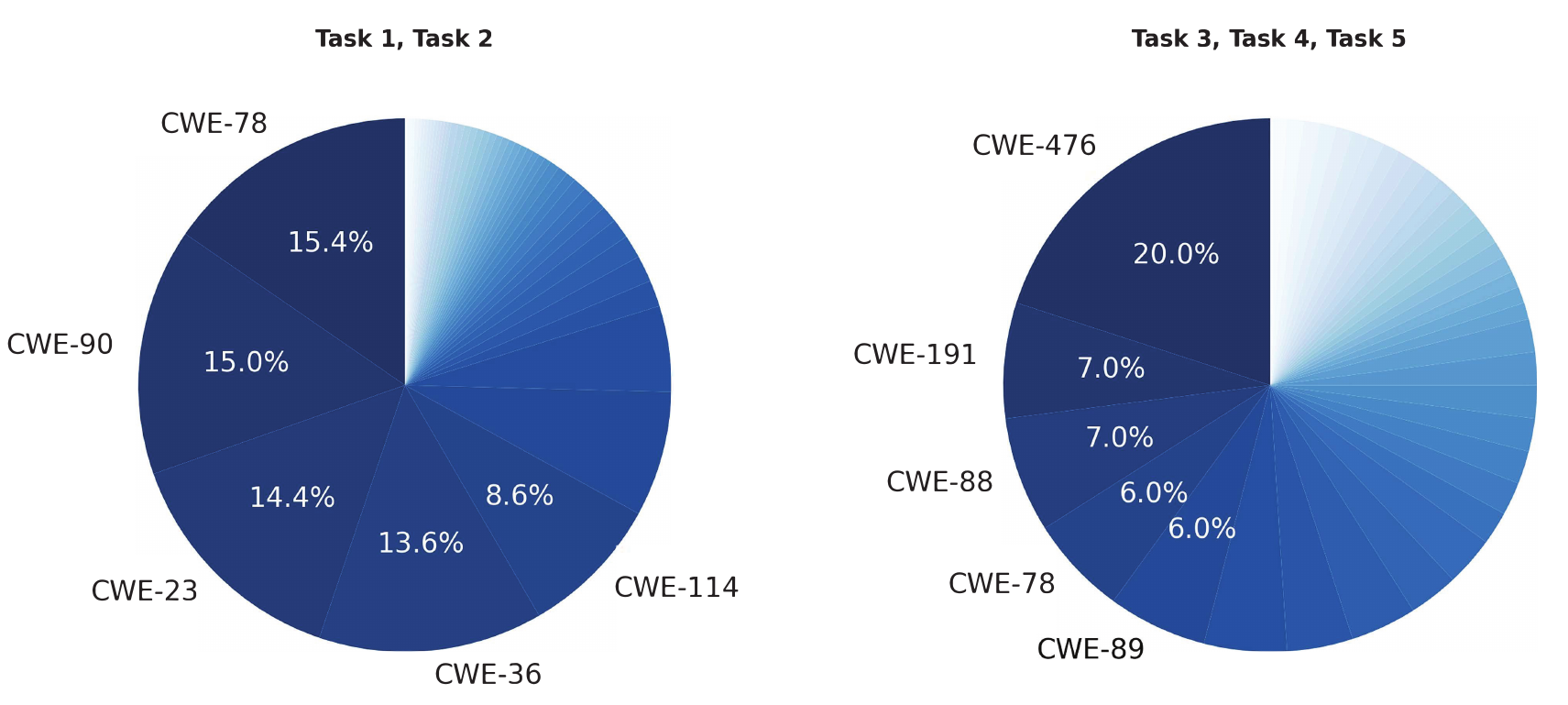}
    \caption{CWE Distribution of VulDetecBench among 5 Tasks}
    \label{fig:cwedistribution}
\end{figure}

In terms of classification, Tasks 1 and 2 feature 48 Common Weakness Enumeration (CWE) classifications, while Tasks 3, 4, and 5 include 38 CWE classifications each. As shown in Figure \ref{fig:cwedistribution}, the distribution of CWE type is designed to be 
reflective of their actual frequency of occurrence in real-world scenarios. This configuration ensures that the results from VulDetectBench closely mimic the performance of LLMs in practical settings. Additionally, for ease of analysis and modification, all benchmark samples are formatted into a ternary structure of \texttt{\{system, user, answer\}}. 

\subsection{Details of Model Selection}

As shown in Table \ref{tab:dom}, we select models with 7B and 13B parameters due to their manageable computational requirements, fitting within the capabilities of most researchers. For our experiments, we standardize resources using NVIDIA GeForce RTX 4090 graphics cards. Models up to 7B parameters run on a single RTX 4090, while larger models (13B and 14B) require two RTX 4090 units in parallel. This setup ensures sufficient computational power and memory, maintaining experimental integrity. For exceptionally large open-source models, we use cloud services accessed via API calls.

\begin{table}[ht]
	\caption{Details Information of LLMs}
	\label{tab:dom}
	\centering
	\resizebox{\linewidth}{!}{
		\begin{tabular}{llcccccc}
			\toprule
			Model Class & Model Version & Size & \makecell{Context \\ Windows}  &  \makecell{Code-specific \\ Pretrain task} & Temp & \makecell{Specifications \\  Computation Requirement} \\
			\midrule
			
			Gemini    & \texttt{gemini-pro}    & -   & 128k   &   & 0 & API                         \\
			Deepseek  & \texttt{deepseek-coder-7b-instruct-v1.5}   & 7B  & 4k   & \Checkmark & 0 & NVIDIA GeForce RTX 4090 * 1 \\
			\multirow{2}*{QWEN}      & \texttt{Qwen1.5-7B-Chat} & 7B  & 32k     &   & 0 & NVIDIA GeForce RTX 4090 * 1 \\
			& \texttt{Qwen1.5-14B-Chat}   & 14B & 32k   &   & 0 & NVIDIA GeForce RTX 4090 * 2 \\
			ChatGLM3  & \texttt{ChatGLM3-6B} & 6B  & 8k  &  &  0 & NVIDIA GeForce RTX 4090 * 1 \\
			\multirow{2}*{Vicuna}    & \texttt{vicuna-13b-v1.5}   & 13B & 4k   &   & 0 & NVIDIA GeForce RTX 4090 * 2 \\
			& \texttt{vicuna-7b-v1.5}  & 7B  & 4k   &   & 0 & NVIDIA GeForce RTX 4090 * 1 \\
			\multirow{2}*{CodeLlama} & \texttt{CodeLlama-13b-Instruct-hf}  & 13B & 16k  & \Checkmark & 0 & NVIDIA GeForce RTX 4090 * 2 \\
			& \texttt{CodeLlama-7b-Instruct-hf }   & 7B  & 16k  & \Checkmark & 0 & NVIDIA GeForce RTX 4090 * 1 \\
			\multirow{4}*{Llama}     & \texttt{Llama3-70B}   & 70B & 8k  &   & 0 & API    \\
			& \texttt{Meta-Llama-3-8B-instruct}  & 8B  & 8k  &   & 0 & NVIDIA GeForce RTX 4090 * 1 \\
			& \texttt{Llama2-7b-chat-hf }  & 7B  & 4k     &   & 0 & NVIDIA GeForce RTX 4090 * 1 \\
			& \texttt{Llama2-13b-chat-hf}   & 13B & 4k   &   & 0 & NVIDIA GeForce RTX 4090 * 2 \\
			\multirow{2}*{Mixtral}     & \texttt{Mixtral-8x7B}   & 7B  & 32k    &  & 0 & API   \\
			& \texttt{Mixtral-8x22B} & 22B & 64k    &   & 0 & API    \\
			Ernie     & \texttt{ernie4.0}   & -   & 8k  &   & 0 & API   \\
			GPT4      & \texttt{gpt4-turbo-0613}   & -   & 128k   &   & 0 & API  \\
			\bottomrule
		\end{tabular}
	}
\end{table}

Most LLMs can now process longer context lengths, but due to dataset quality and inference costs, we limit VulDetectBench's input context to under 4K tokens with each case in dataset including only one vulnerability. Longer contexts risk including multiple vulnerabilities in a single sample. However, as models evolve and support longer window size, we have designed a dataset with longer code length for evaluation including 8k, 16k, 32k, etc, and will update it to the repository.

\subsection{Details of Evaluation Process}

\begin{figure}[htb]
    \centering
    \includegraphics[width=1\linewidth]{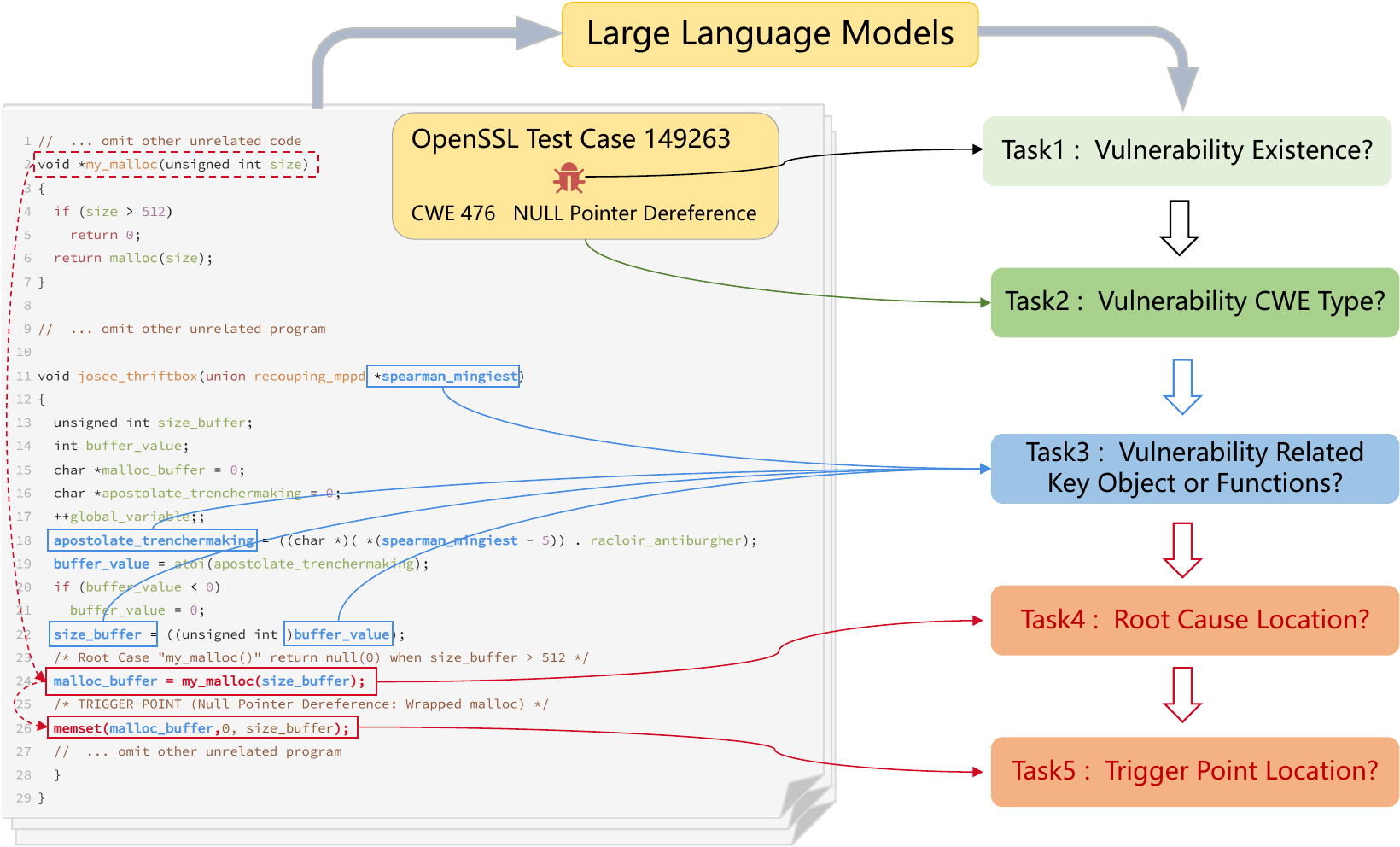}
    \caption{Capability Evaluation of LLM on Vulnerability Detection. The figure shows the specific design of the five tasks in VulDetechBench. Tasks 1 and 2 are classification tasks where the LLM is asked to determine whether a vulnerability exists in a program and its CWE type. Task 3 evaluates the key data object and function call associated with the vulnerability, as shown in the blue box in the figure. Tasks 4 and 5 are more difficult and require the LLM to locate the root cause of the vulnerability and the vulnerability's trigger point.}
    \label{fig:pipeline}
\end{figure}

This section details the evaluation process undertaken by VulDetectBench on the program, as depicted in Figure \ref{fig:pipeline}. Tasks 1 and 2 focus on identifying the presence of vulnerabilities in a program and classifying their specific types. These tasks assess the  capability of large language models (LLMs) to comprehend the overall vulnerability landscape within a program. In the main text, we thoroughly discuss whether the outcomes of Tasks 1 and 2 accurately represent the LLM's specific ability to understand vulnerabilities. 

Task 3 focuses on the Large Language Model's (LLM) capability to identify key data objects or function calls within the program, elements that are critically relevant to vulnerabilities. Tasks 4 and 5 extend this examination, where the program's root cause and trigger points are typically associated with specific lines of code. While the number of lines related to the root cause might be extensive due to the inherent complexity of the program, these lines nonetheless represent only a small fraction of the entire codebase. 

This variance in the number of lines that correspond to the root causes and trigger points underscores the challenges posed by the program's complexity. It necessitates advanced analytical capabilities from the LLMs, highlighting the high demand placed on these models to precisely identify and isolate critical segments within vast and intricate programming environments. This complexity elevates the LLM's operational requirements.

\subsection{Comparision with Traditional Tools}
 Traditional static program analysis tools such as Infer require compilation of the program, which transforms program syntax trees (ASTs), program control flow graphs (CFGs), and other forms of program representation. CodeQL requires the use of manually constructed databases to perform vulnerability matching. This does not allow for rapid vulnerability analysis, and we choose a commonly used tool  flawfinder that can directly analyze portions of a program's code for comparative experiments. This tool can complete five tasks in VulDetectBench, but none of them give correct results and produce a large number of false positives. 
\begin{table}[htbp]
\centering

\caption{Analysis Result of Vulnerability from Tradition Tools : Flawfinder}
\resizebox{\linewidth}{!}{
    \begin{tabular}{ccccccl}
    
    \toprule
    Line & Level & Category & Name & CWEs & Context \\
    \midrule
    46  & 4 & buffer & sprintf & CWE-120 & \texttt{sprintf(dirpath, "\%s/\%s", ss\_tc\_root, "testData")} \\
    55  & 4 & buffer & sprintf & CWE-120 & \texttt{sprintf(filepath, "\%s/\%s", dirpath, "logfile.txt")} \\
    67  & 4 & format & printf & CWE-134 & \texttt{void printf(char *format, ...)\{} \\
    70  & 4 & format & vfprintf & CWE-134 & \texttt{vfprintf(printf\_context, format, argptr);} \\
    41  & 3 & buffer & getenv & CWE-807, CWE-20 & \texttt{ss\_tc\_root = getenv("SS\_TC\_ROOT");} \\
    211 & 3 & buffer & getenv & CWE-807, CWE-20 & \texttt{humongous\_phocomelous = getenv("UNAPPROACHABLY\_MYRIAPODAN");} \\
    56  & 2 & misc   & fopen   & CWE-362 & \texttt{printf\_context = fopen(filepath, "w");} \\
    90  & 2 & misc   & fopen   & CWE-362 & \texttt{file = fopen(filename,mode);} \\
    206 & 2 & buffer & char    & CWE-119/CWE-120 & \texttt{char p[4];} \\
    319 & 2 & misc   & fopen   & CWE-362 & \texttt{fp = fopen(ptr,p);} \\
    442 & 2 & buffer & char    & CWE-119/CWE-120 & \texttt{char stack\_string[stack\_size];} \\
    \bottomrule
    \end{tabular}
    \label{tab:traditionaltool}
}

\end{table}

In our study, we analyze Test Case 149263 in OpenSSL, which contains a Null Pointer Dereference vulnerability classified under CWE-476. As shown in Table \ref{tab:traditionaltool}, the analysis is based on the results from Flawfinder's evaluation of this case. Flawfinder fails to detect the CWE-476 type of vulnerability and generates a significant number of false positives. Additionally, it is unable to identify the key data objects associated with the vulnerability. These results prove insufficient for providing valuable insights for security analysis.

For static testing tools, like Flawfinder presented, it just treats the program as text and uses fixed patterns to detect and give a report of vulnerability. And we can see it makes lots of false positives and hardly provides useful information for security analysts. So, to conduct a more precise program analysis, like inter-procedure program analysis and context-sensitive analysis and so on, the program need to be transfered into AST(abstract syntax trees) or some IR representaions(like LLVM IR) like CodeQL \cite{githubCodeQL} and Cppcheck\cite{sourceforgeCppcheckTool} . That requires the program are complete and can be successfully compiled.

\subsection{ Deep Understanding of the Capability of Vulnerability Analysis}

In this section,we provide detailed data about the  analysis of models from the same family but of varying sizeson their performance across the aforementioned five tasks. 
\begin{table}[htb]
	\centering
	\caption{Performance comparison of different sizes within the same LLM family}
	\resizebox{\linewidth}{!}{
		\begin{tabular}{lcccccc}
			\toprule
			Model  &  Size  & \makecell{Task1 \\ ACC   |    F1 } &  \makecell{Task2  \\   SE | ME} &   \makecell{Task 3   \\ MAR|   MIR }&  \makecell{Task 4  \\  URS  | ORS  } & \makecell{Task 5  \\ URS | ORS   } \\
			\midrule
		\multirow{2}*{Mixtral}  & 8*7B                  & 76.42 |   79.00          & 62.00 |   58.40                         & 21.61 |   17.51 & 01.66 |   04.28          & 05.74 |   03.51          \\
			 & 8 *22B                 & 81.82 |   84.47          & 77.80 |   74.80                       &  30.26  |   24.10 & 17.49 |   11.46          & 06.17 |   02.83     \\  
    \midrule
   \multirow{2}*{Llama2} & 7B              & 47.90 |   64.19          & 41.40 |   54.40                         & 11.47 |   09.59 & 0 |   0                 & 0 |   0                \\
			 & 13B           & 70.37 |   73.67          & 58.70 |   67.40                         & 10.99 |   09.23 & 00.50 |   01.20           & 00.10 |   00.20  \\
    \midrule
  
		 \multirow{2}*{CodeLlama} & 7B        & 36.40 |   53.37          & 36.70 |   41.60                         & 06.28 |   05.04& 01.69 |   02.55          & 00.69 |   01.31  
   \\
			 & 13B      & 47.90 |   58.81          & 32.80 |   44.60                         & 10.34 |   08.51 & 03.30 |   03.47          & 01.29 |   01.89          \\
			
	\midrule

			\multirow{2}*{Qwen} & 7B                 & 62.30 |   43.31          & 60.10 |   62.20                         & 13.63 |   11.15 & 04.34 |   10.09          & 05.56 |   10.95          \\
			 & 14B              & 69.10 |   55.67          & 77.90 |   82.80                         & 16.49 |   13.67 & 06.40 |   08.32          & 03.14 |   04.81 
  \\

			\bottomrule
		\end{tabular}
		\label{tab:performancecompare}
	}
	
\end{table}

To better understand the extent of the LLMs' understanding of vulnerabilities, we use 100 identical test cases from tasks one to five for analysis. In the performance comparison experiments, shown in Table \ref{tab:100direct}, we evaluate the performance of the model on 100 samples of different tasks. We chose the five LLMs with the best overall capabilities. Here we unify the metrics of each task into a score out of 100 in order to reflect the decrease in model capability. The score for task 1 is from ACC, the score for task 2 is from SE the score for task 3 is from MAR, and the scores for task 4 and task 5 are from ORS. Meanwhile, Table \ref{tab:100indirect} shows the performance of the model when the previous task was executed correctly (the number in task 2 is the number that was correct in task 1). As a whole, LLMs are unable to accurately identify data objects and critical function calls that are closely associated with vulnerabilities, nor can they accurately locate root causes and trigger points. 



\begin{table}[ht]
\centering
\caption{Performance on 100 samples of different
tasks.}
\begin{tabular}{llllll}
\toprule
Model             & Task1 & Task2 & Task3  & Task4 & Task5 \\
\midrule
GPT4              & 100.0  & 76.5  & 32.5 & 27.1 & 17.9 \\
Gemini-Pro        & 58.0    & 51.0    & 17.2  & 23.4 & 18.9 \\
Qwen1.5-14B       & 60.0    & 57.5  & 26.9  & 8.3  & 4.8   \\
Meta-Llama-8B     & 100.0   & 52.5  & 36.5   & 0.5   & 1.1   \\
Deepseek-coder-7B & 70.0    & 16.0   & 28.6  & 9.2   & 8.8  
\\
\bottomrule
\end{tabular}

\label{tab:100direct}
\end{table}

\begin{table}[ht]
\centering
\caption{Performance of the model given that the previous task was performed correctly.For example, if there are 80 correct answers in Task 1, see how many of those 80 answers are correct in Task 2, and then assess Tasks 3,4,5 in the cases where Task 2 is answered correctly.}
\begin{tabular}{llllll}
\toprule
Model             & Task1 & Task2 & Task3  & Task4 & Task5 \\
\midrule
GPT4              & 100.0   & 76.5  & 28.7 & 22.6 & 14.8  \\
Gemini-Pro        & 58.0    & 35.0    & 8.4   & 6.0     & 6.0     \\
Qwen1.5-14B       & 60.0    & 31.5  & 8.8   & 2.9   & 1.0     \\
Meta-Llama-8B     & 100.0  & 52.5  & 24.3  & 0.3   & 1.1   \\
Deepseek-coder-7B & 70.0    & 10.5  & 4.7   & 2.3   & 1.5  \\
\bottomrule
\end{tabular}

\label{tab:100indirect}
\end{table}

Despite their ability to accurately determine the presence of vulnerabilities and even identify CWE types, this does not mean that the models truly understand the specific details of the vulnerabilities.  If we divide the five tasks into side-by-side tasks, although the performance of LLM decreases on the more difficult tasks, it still fails to reflect the model's ability to analyse each vulnerability case in-depth, and we can see that the LLM model's ability to analyse each specific vulnerability case decreases even more severely as shown in Table \ref{tab:100indirect}.This suggests that large models assess the presence of vulnerabilities primarily from the perspective of high-level language features, and lack a deeper understanding of the specific mechanisms of vulnerability occurrence.

\subsection{Finetuing the LLM}
At the same time, we have carried out finetuning experiments, which mainly focused on tasks 3, 4 and 5, which had poor performance of the fundamental model. Task 3 was the intermediate analysis process of tasks 4 and 5, so we could directly fine-tune tasks 4 and 5 without considering task 3. Through the analysis of the experiment of fine-tuning Meta-Lema-3-8B-Instruct, we found that the fundamental model basically did not show any ability for the two tasks. 
The model used in this evaluation is the Meta-Llama-3-8B-Instruct, which is designed for the CAUSAL\_LM task type. The rank is set to 8. The model is fine-tuned using a LoRA (Low-Rank Adaptation) with an alpha value of 32 and a dropout rate of 0.1. The learning rate applied during fine-tuning is 2e-4, and the model is trained over 5 epochs.
\begin{table}[h!]
\centering
\caption{Performance of Fine-tuned Models on TASK4 and TASK5}
\begin{tabular}{lcc}
\hline
\textbf{Model} & \textbf{TASK4} & \textbf{TASK5} \\ \hline
Meta-Llama-3-8B-Instruct & 01.10 & 00.53 \\   
Meta-Llama-3-8B-Instruct (Fine-tuned) & 39.00 & 52.07 \\ \hline
\end{tabular}

\label{tab:task_performance}
\end{table}

After fine-tuning, the recall rate increased to 39.00\% for task4, and 52.07\% for task5. we can see that there is a definite improvement. However, due to strict evaluation requirements, the effect is only improved to a certain extent under the current metric. We conducted an in-depth analysis of the model's responses and found that fine-tuning significantly enhanced its ability to perceive vulnerability locations. However, pinpointing a specific line remains a challenging task, and we are continuing to work on it. 

\subsection{Chain of Thought and Few-shot Testing}
Furthermore, we conducted additional experiments to investigate the model's performance on vulnerability detection tasks under different prompting technique settings. We used GPT-4 and tested it on tasks 3, 4, and 5 using both Chain of Thought (CoT) and Few-shot prompting techniques. The results are as follows.

\begin{table}[h!]
\centering
\caption{Experimental results for vulnerability detection tasks using CoT.}
\begin{tabular}{lccc}
\hline
\textbf{Method} & \textbf{TASK3} & \textbf{TASK4} & \textbf{TASK5} \\ \hline
Zero-shot & 20.21|16.07 & 24.26|27.07 & 13.00|17.85 \\ 
CoT & 45.54|39.85 & 1.25|16.77 & 1.64|22.88 \\ 
Few-shot (2-shot) & 10.34|13.41 & 9.64|9.64 & 15.37|15.37 \\ \hline
\end{tabular}

\label{tab:results}
\end{table}

Experimental results show that CoT improves vulnerability analysis, while few-shot prompting may negatively impact localization. We attribute this to the following reasons:
\begin{enumerate}
    \item Vulnerability diversity makes few-shot examples potentially misleading.
    \item CoT, designed based on expert analysis methods, aligns with standard vulnerability assessment paradigms, improving performance in tasks 3 and 5.
\end{enumerate}

\subsection{Prompt Design}

We provided five task-specific prompts, and we tested the quality of each prompt, both in terms of whether it produces the desired output and whether it creates ambiguity that affects the model's answers.

In Tasks 1 and 2, we require the model to provide specific outputs for the final evaluation to test the stability of prompts. We use the same prompt for each of the selected models across 10 times and recorded the number of times we obtained the desired result. Obtaining the desired result involves receiving outputs from the model that can be directly used for metric evaluation, such as a consistent "Yes" or a singular selected option. As shown in Table \ref{tab:pqt}, LLMs can output desired results for each times prompting. This method ensures that we can reliably assess the prompt's effectiveness and the model's consistency in producing expected outcomes. 
\begin{table}[h]
    \centering
    \caption{Prompt Quality Testing. }
    \resizebox{\linewidth}{!}{
        \begin{tabular}{c|c|c|c|c|c|c}
        \toprule
        \textbf{} & \textbf{CodeLlama-7B} & \textbf{CodeLlama-13B} & \textbf{Llama2-7B} & \textbf{Llama2-13B} & \textbf{Gemini-Pro} & \textbf{Mistral-7B} \\ 
        \midrule
        \textbf{Expected Answer Task1} & 10/10                 & 10/10                  & 10/10              & 9/10                & 10/10              & 8/10              \\ 
        \textbf{Expected Answer Task2} & 10/10                 & 10/10                  & 10/10              & 10/10                & 10/10              & 8/10              \\ 
        \bottomrule
    \end{tabular}
    }
    \label{tab:pqt}
\end{table}

For Tasks 4 and 5, we also evaluate whether the terms “root cause” and “trigger point” would affect the model's understanding of the task, and we compare the model outputs corresponding to the prompt with the relevant explanations added and the prompt with only the terms and find that the model has the ability to understand the aforementioned terms, and would not affect the model's understanding of the task.

\begin{figure}
    \centering
    \includegraphics[width=1.0\linewidth]{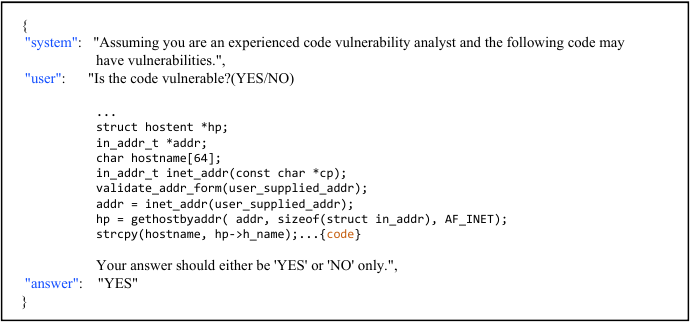}
    \caption{Sample for Task 1: Vulnerability Existence Detection}
    \label{fig:prompts1}
\end{figure}

\begin{figure}
    \centering
    \includegraphics[width=1.0\textwidth]{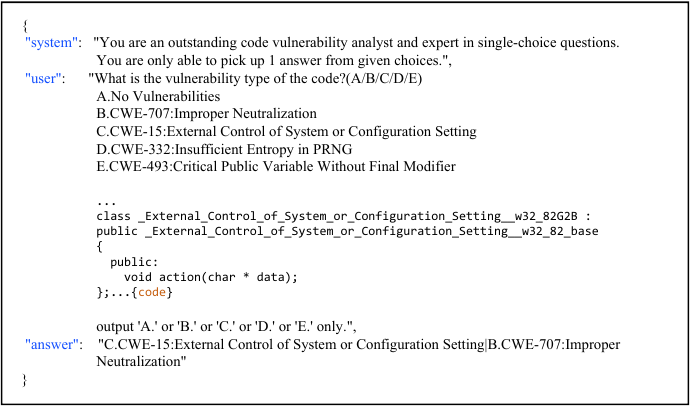}
    \caption{Sample for Task 2: CWE Type Inference}
    \label{fig:prompts2}
\end{figure}

\begin{figure}
    \centering
     \includegraphics[width=\textwidth]{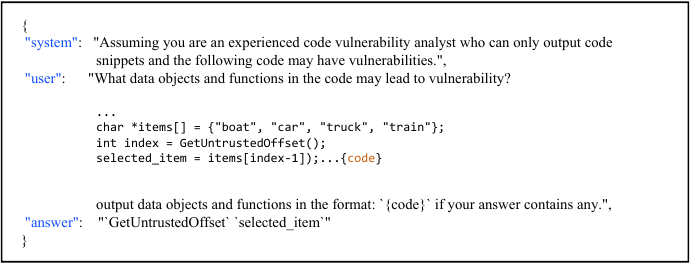}
    \caption{Sample for Task 3: Key Data Objects and Functions Identification}
    \label{fig:prompts3}
\end{figure}

\begin{figure}
    \centering
    \includegraphics[width=\textwidth]{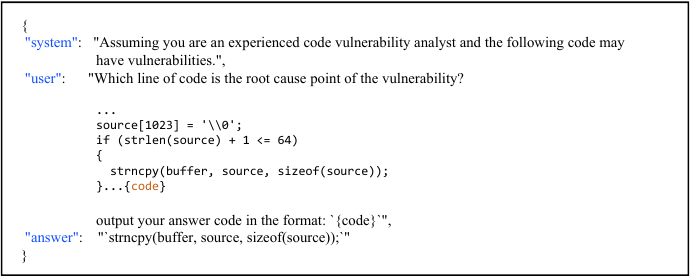}
    \caption{Sample for Task 4: Root Cause Location}
    \label{fig:prompts4}
\end{figure}

\begin{figure}
    \centering
    \includegraphics[width=\textwidth]{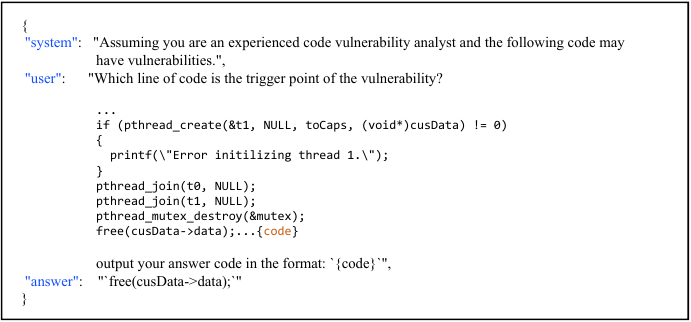}
    \caption{Sample for Task 5: Trigger Point Location}
    \label{fig:prompts5}
\end{figure}

The tasks are exampled by Figure  \ref{fig:prompts1},\ref{fig:prompts2},\ref{fig:prompts3},\ref{fig:prompts4} and \ref{fig:prompts5}. In each sample, "system" and "user" are input to the model as system and user prompts, respectively."answer" is the expected output of the model.

\clearpage


\end{document}